\documentclass[twocolumn,aip,amsmath,amssymb,reprint,citeautoscript]{revtex4-1}  % for review and submission
\usepackage[final]{pdfpages}
\usepackage{graphicx}  % needed for figures
\usepackage{dcolumn}   % needed for some tables
\usepackage{bm}        % for math
\usepackage{amssymb}   % for math
\usepackage{hyperref}
\usepackage{amsmath}
\usepackage{mathtools}
\usepackage{float}
\usepackage{color}

\allowdisplaybreaks

\makeatletter
\AtBeginDocument{\let\LS@rot\@undefined}
\makeatother

\newcommand{\citeasnoun}[1]{Ref.~\citenum{#1}}
\renewcommand{\vec}[1]{\mathbf{#1}}
\newcommand{\figref}[1]{Fig.~\ref{#1}}
\newcommand{\Figref}[1]{Figure~\ref{#1}}
\newcommand{\re}[1]{\text{Re}\left(#1\right)}

\begin{document}

\title{Is single-mode lasing possible in an infinite periodic system?}

\author{Mohammed Benzaouia} \email{medbenz@mit.edu} \affiliation{Department of Electrical Engineering and Computer Science, Massachusetts Institute of Technology, Cambridge, MA 02139, USA.} 
\author{Alexander  Cerjan} \affiliation{Department of Physics, Penn State University, State College, PA 16801, USA.}
\author{Steven G. Johnson} \affiliation{Department of Mathematics, Massachusetts Institute of Technology, Cambridge, MA 02139, USA.}

\begin{abstract}
In this Letter, we present a rigorous method to study the stability of periodic lasing systems. In a linear model, the presence of a continuum of modes (with arbitrarily close lasing thresholds) gives the impression that stable single-mode lasing cannot be maintained in the limit of an infinite system. However, we show that nonlinear effects of the Maxwell--Bloch equations can lead to stable systems near threshold given a simple stability condition on the sign of the laser detuning compared to the band curvature. We examine band-edge (1d) and bound-in-continuum (2d) lasing modes and validate our stability results against time-domain simulations.   
\end{abstract}
\maketitle

\begin{figure}
\includegraphics[width=\columnwidth, keepaspectratio]{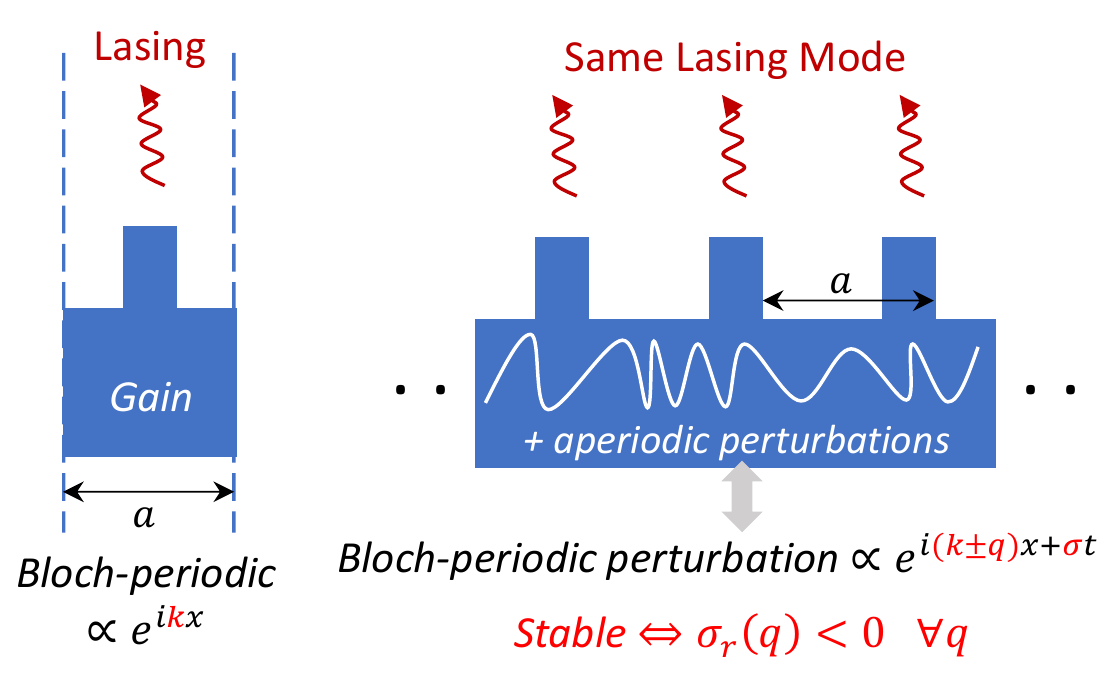}
\caption{We study the stability of a single Bloch-periodic lasing mode under \emph{aperiodic} perturbations. The stability eigenproblem can be solved using Bloch theorem by writing perturbations as a general Bloch wave. The lasing mode is stable when real parts of the eigenvalues $\sigma(q)$ are negative for all wavevectors $q$.}
\label{Fig1}
\end{figure}

Many lasers rely on resonances in periodic systems, ranging from band-edge modes of grated distributed-feedback (DFB) waveguides~\cite{kogelnik1971stimulated, carroll1998distributed} or photonic-crystal surface-emitting laser (PCSELS)~\cite{imada1999coherent, meier1999laser, noda2001polarization, kurosaka2010chip, zhou2013lasing, hirose2014watt, zhao2016printed} to more exotic bound-in-continuum (BiC) states~\cite{kodigala2017lasing, ha2018directional}.   In this Letter, we address a fundamental question for periodic lasers: does stable single-mode lasing exist in an infinite periodic structure, or does it inherently require the boundaries of a finite structure to stabilize?   A number of theoretical works have studied lasing with periodic boundary conditions as in \figref{Fig1}(left) and found lasing modes~\cite{chua2011low, wuestner2010overcoming, marani2012gain, dridi2013model, cuerda2015theory, droulias2017novel}, but neglected a key concern: even if the structure and the lasing mode are periodic, stable lasing requires that arbitrary \emph{aperiodic} electromagnetic perturbations [as in \figref{Fig1}(right)] must decay rather than grow~\cite{glendinning1994stability, burkhardt2015steady, liu2017symmetry}. At first glance, such stability may seem unlikely: any resonance in a periodic system is part of a \emph{continuum} of resonances at different Bloch wavevectors with arbitrarily close lasing thresholds, and this seems to violate typical assumptions for stable lasing~\cite{tureci2006self, ge2008quantitative, esterhazy2014scalable}. A finite-size structure discretizes the resonance spectrum and hence may suppress this problem, but instabilities have been observed in large enough finite periodic lasers where the resonances become very closely spaced~\cite{liang2014mode}. Analogous transverse instabilities are known to occur in translation-invariant ($\mathrm{period}\to 0$) lasers such as VCSELs~\cite{iga2003vertical}, for which stability analysis has been performed with various assumptions~\cite{san1995light, mandel2004transverse}. In fact, however, we show that single-mode lasing \emph{is} possible even in \emph{infinite} periodic structures for range of powers above threshold, by applying a Bloch adaptation of linear-stability analysis to the full Maxwell--Bloch equations~\cite{burkhardt2015steady, liu2017symmetry}. (Instabilities can still arise if our criteria are violated, or from effects such as disorder not considered in this work.) We consider examples for both 1d DFB-like lasers and 2d BiC-based lasing~\cite{hsu2013bloch, kodigala2017lasing, ha2018directional}, and validate our result against brute-force time-domain simulations~\cite{OskooiRo10, 2020arXiv}. Using perturbation theory (in the supplementary material), we also obtain a simple condition for stability near threshold of low-loss resonances and confirm it numerically: the sign of the laser detuning from the gain frequency should match the sign of the band curvature at threshold.  

\medskip

We consider lasing systems described by the semi-classical Maxwell--Bloch equations (with the rotating-wave approximation), which fully include nonlinear mode-competitition effects (such as spatial hole-burning)~\cite{haken1986laser}:  
\begin{align*}
-\nabla \times \nabla \times \mathbf{E}^+ &=  \mathbf{\ddot P^+} + \epsilon_c\mathbf{\ddot E^+} +\sigma_c \mathbf{\dot E^+} \\
i\mathbf{\dot P^+} &= (\omega_a-i\gamma_\perp)\mathbf{P}^+ + \gamma_\perp \mathbf{E}^+D \stepcounter{equation}\tag{\theequation}\label{MB} \\
\dot D/\gamma_\parallel &= D_0-D+\text{Im} (\mathbf{E^{+*}}\cdot \mathbf{P}^+),
\end{align*}
where $\mathbf{E}^+$ is the positive-frequency component of the electric field (the physical field being given by $2\text{Re}[\mathbf{E}^+]$), $\mathbf{P}^+$ is the positive-frequency polarization describing the transition between two atomic energy levels (with frequency $\omega_a$ and linewidth $\gamma_\perp$), $D$ is the population inversion (with relaxation rate $\gamma_\parallel$), $D_0$ is the pump strength, $\epsilon_c$ is the cold-cavity real permittivity, and $\sigma_c$ is a cold-cavity conductivity loss. Here, we are assuming that the orientation of the atomic transition is parallel to the electric field, and have written all three fields in their natural units~\cite{burkhardt2015steady}.

A steady-state solution of these equations can be obtained via steady-state ab-initio lasing theory (SALT), which is exact for single-mode lasing and approximate for multi-mode lasing with well-separated modes~\cite{tureci2006self,ge2008quantitative,esterhazy2014scalable}.  For a periodic system, we consider a Bloch-mode steady-state solution $\mathbf{E}^+ =  \mathbf{E_\vec{k}}e^{i(\vec{k}\cdot \mathbf{x}-\omega t)}$ satisfying the stationary ($\dot D=0$) SALT equation:
\begin{equation} \label{steady-state}
\Theta_\vec{k} \mathbf{E_\vec{k}} = \omega_\vec{k}^2\left[ \epsilon_c+i\frac{\sigma_c}{\omega_\vec{k}} +\Gamma(\omega_\vec{k})D_\vec{k} \right] \vec{E}_\vec{k},
\end{equation}
where $\Gamma(\omega) = \gamma_\perp/\left(\omega-\omega_a+i\gamma_\perp \right)$, $\vec{P}_\vec{k} = \Gamma(\omega_\vec{k})D_\vec{k} \vec{E}_\vec{k}$, $D_\vec{k} = D_0 / \left(1 + |\Gamma(\omega_\vec{k}) \vec{E}_\vec{k}|^2\right)$ and $\Theta_\vec{k} = e^{-i\vec{k}\cdot \mathbf{x}} \nabla \times \nabla \times e^{i\vec{k}\cdot \mathbf{x}}$ is a periodic operator.

Given this steady-state solution, one can then apply linear-stability analysis to the full Maxwell--Bloch equations, linearizing \emph{arbitrary aperiodic} perturbations $X= X_\vec{k} + \delta X$, for $X \in \{\vec{E},\vec{P},D\}$, to determine whether perturbations $\delta X$ exponentially grow (unstable) or shrink (stable)~\cite{glendinning1994stability, burkhardt2015steady, liu2017symmetry}.   Here, our key point is that, because the linearized equations for the perturbations $\delta X$ are periodic (for a Bloch-mode steady state), we can apply Bloch's theorem~\cite{tinkham2003group} to decompose the perturbations \emph{themselves} into Bloch-wave modes $\delta \vec{E}_{\vec{q}}$, solving a separate linear-stability eigenproblem for each wavevector~$\vec{q}$.  

The well-known linear-stability analysis~\cite{burkhardt2015steady} of the Maxwell--Bloch equations~(\ref{MB}) proceeds as follows. Linearization of~(\ref{MB}) in $\delta X$ gives:
 \begin{equation}\label{lin-MB} \begin{split}
\mathbf{0} & =  \Theta_\vec{k} \delta \mathbf{E} +d_\omega^2(\epsilon_c \delta \mathbf{E}+\delta \mathbf{P}) +d_\omega\sigma_c \delta \mathbf{E} \\
 i \delta \mathbf{\dot{P}} & = (\omega_a-\omega-i\gamma_\perp)\delta \mathbf{P} + \gamma_\perp (D_\vec{k}\delta \mathbf{E}+\mathbf{E_\vec{k}}\delta D)\\
 \delta \dot D/ \gamma_\parallel & = - \delta D + \text{Im}(\mathbf{P_\vec{k}}\cdot \delta \mathbf{E}^*+\mathbf{E}_\vec{k}^*\cdot \delta \mathbf{P}),
\end{split} \end{equation}
where $d_\omega =  \left(\frac{d}{dt}-i\omega \right)$. Splitting complex variables into real and imaginary parts yields a set of linear equations $\left(C\frac{d^2}{dt^2}+B\frac{d}{dt}+A \right)u(\vec{x},t) = 0$~\cite{burkhardt2015steady}, where $u = (\text{Re}(\delta \mathbf{E}), \text{Im}(\delta \mathbf{E}), \text{Re}(\delta \mathbf{P}), \text{Im}(\delta \mathbf{P}), \delta D )$ and $A$, $B$ and $C$ are operator matrices readily obtained from~(\ref{lin-MB}).  Stability analysis consists of looking for solutions of the form $u=\text{Re}(U e^{\sigma t})$, which leads to a quadratic eigenproblem: 
 \begin{equation}\label{instab-eigen}\left( A+B\sigma + C\sigma^2 \right)U = 0. \end{equation}  
 The sign of $\text{Re}(\sigma)$ determines the stability of the single-mode solution~\cite{burkhardt2015steady}.

Since the operators $A$, $B$ and $C$ are periodic in our case, however, we can use Bloch's theorem to further simplify the problem: the eigenfunctions can be chosen in the Bloch form $U=U_{\vec{q}} e^{i\vec{q}\cdot \mathbf{x}}$ where $U_{\vec{q}}$ is periodic. The eigenvalues $\sigma(\vec{q},D_0)$ then determine the stability: If there exists a wavevector $\vec{q}$ so that $\text{Re}(\sigma(\vec{q},D_0))>0$, then the single-mode solution is unstable at the pump rate $D_0$, with exponential growth at the wavevector $\vec{k}\pm\vec{q}$.  Since $(A, B, C)$ are real, we also have $\sigma(\vec{q},D_0)=\sigma(-\vec{q},D_0)^*$, so we need only consider one side of $\vec{q}$ within the Brillouin zone.

\begin{figure}
\includegraphics[width=\columnwidth, keepaspectratio]{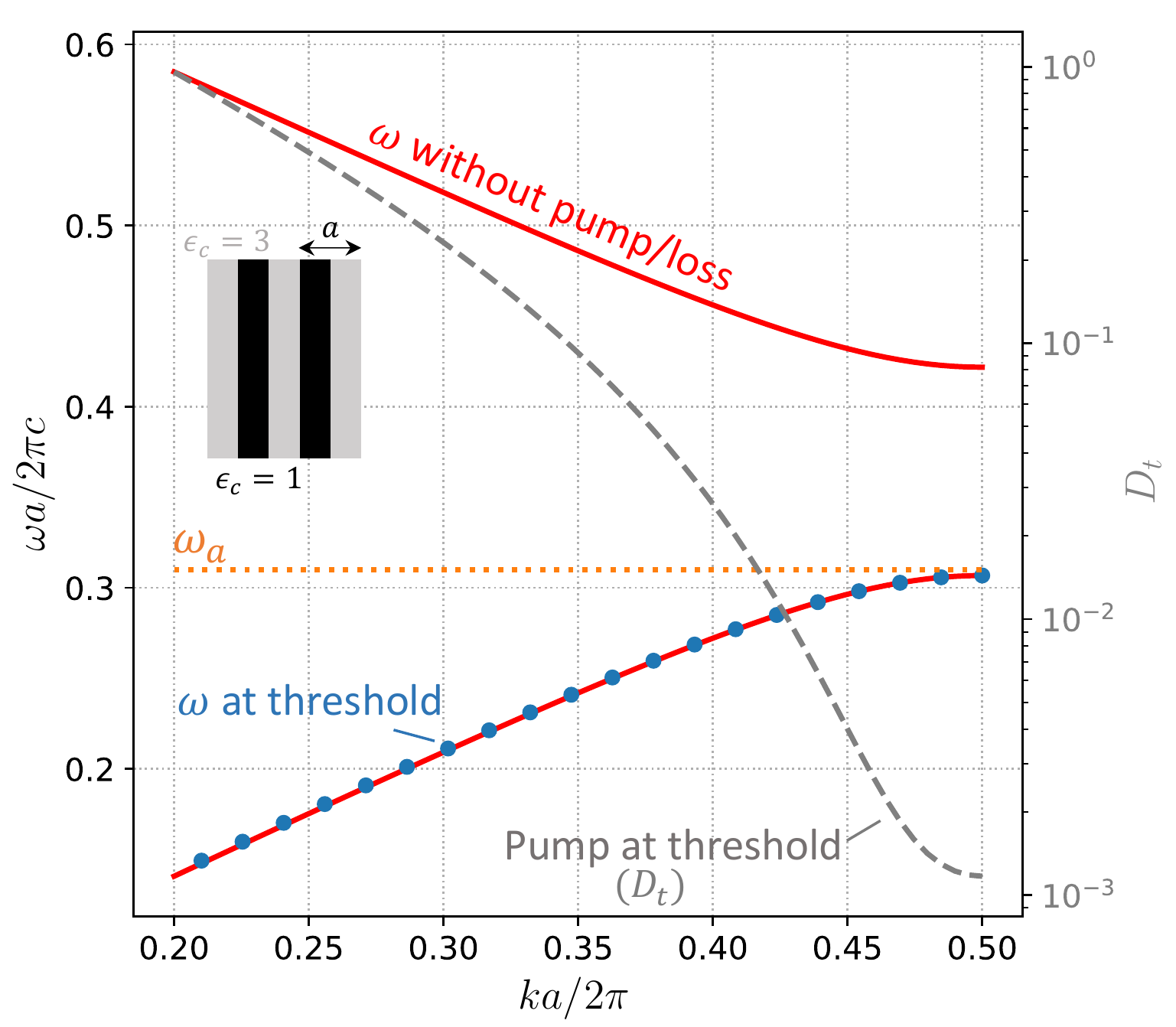}
\caption{The cold cavity is 1D photonic crystal with uniform conductivity loss $\sigma_c=0.001\omega_a$. The two-level gain medium is characterized by $\omega_aa/2\pi c=0.31$ and $\gamma_\perp a/2\pi c=0.008$. The frequency (dots) and pump (dashed lines) at the lasing threshold are computed for modes of the first band. The minimum pump at threshold is obtained at the band edge $ka=\pi$. In absence of gain, the decay rate for the band-edge mode is equal to $\kappa \approx 5.8 \times 10^{-5} (2\pi c /a)$. }
\label{Fig2}
\end{figure}

\medskip

We can now use this method to study a simplified model for a DFB laser formed by a 1D photonic crystal with alternating layers of equal thickness and dielectric constants equal to 1 and 3 (Figure \ref{Fig2}). We assume a uniform conductivity loss $\sigma_c=0.001\omega_a$ and a two-level gain medium with $\omega_aa/2\pi c=0.31$ and $\gamma_\perp a/2\pi c=0.008$. \Figref{Fig2} shows part of the band diagram, with $\omega_a$ chosen near the first band edge. For every wavevector $k$ of the first band, we compute the pump threshold $D_t$, defined as the lowest pump rate $D_0$ that compensates the loss and leads to a \emph{real} eigenfrequency $\omega_k$ in~(\ref{steady-state}). As expected, the smallest $D_t$ is obtained at the band edge $k=\pi/a$ of the first band, which we therefore take to be the first lasing mode. However, as discussed earlier, $D_t$ varies continuously with $k$ and other modes are expected to reach threshold for arbitrary close values of the pump in the \emph{linear} model. 
 
\begin{figure*}
\includegraphics[width=2\columnwidth, keepaspectratio]{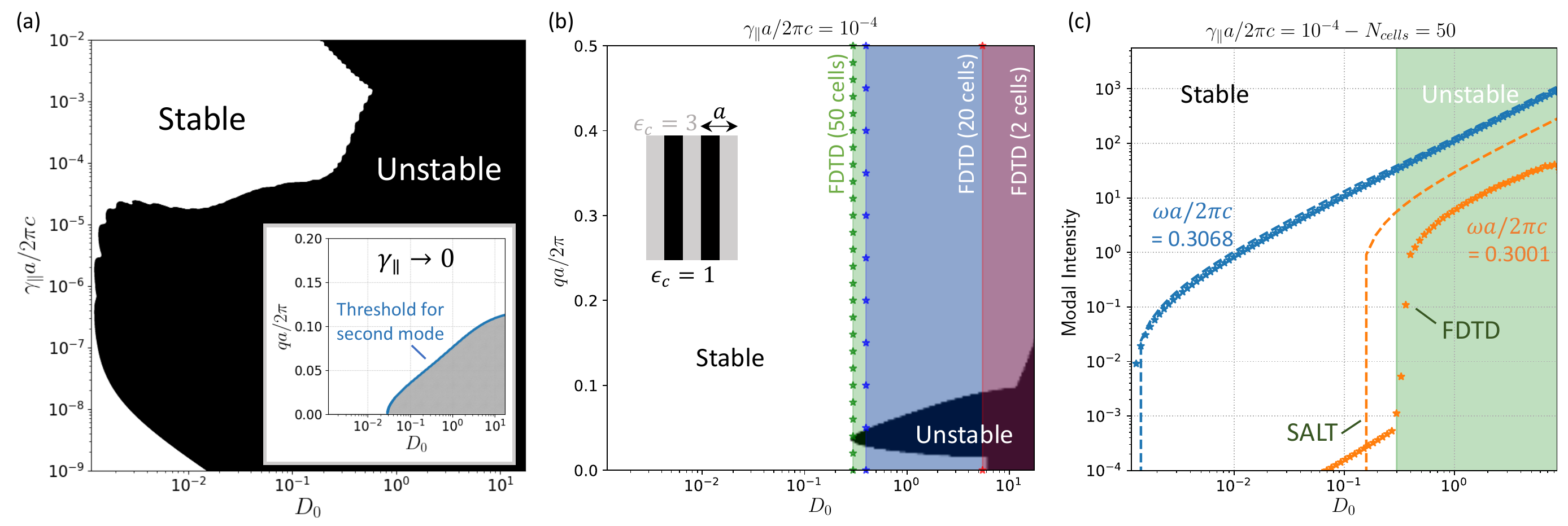}
\caption{(a) Stability region obtained from Maxwell--Bloch stability eigenproblem as a function of $\gamma_\parallel$ and pump strength $D_0$. Inset shows the pump threshold of the second lasing mode using multimode SALT (assuming one first mode at $ka=\pi$ is lasing). This represents the limit $\gamma_\parallel \rightarrow 0$ of the stability eigenproblem. (b) Detailed stability map for $\gamma_\parallel a/2\pi c=10^{-4}$ as a function of $q$. We compare results to FDTD simulations using a finite supercell with periodic boundary conditions (unstable in shaded regions), initialized with the SALT solution plus $\sim 1\%$ noise and checking stability after $\sim 10^5$ optical periods. Stars show the allowed $q$ due to the finite supercell ($2\pi \ell/aN_\mathrm{cells}$). (c) Modal intensity of lasing modes with FDTD ($N_\mathrm{cells}=50$) and multimode SALT (assuming second lasing mode at $q=4\pi/50a$).}
\label{Fig3}
\end{figure*}

In order to study the stability of the lasing band-edge mode, we first solve the steady-state nonlinear equation~(\ref{steady-state}) at higher pump values with a Newton-Raphson solver as described in \citeasnoun{esterhazy2014scalable}. We then use the obtained steady-state solution to solve the stability eigenproblem~(\ref{instab-eigen}) for different pump values. Results are summarized in \figref{Fig3}. First, note that the single mode solution is \emph{stable} close to threshold, unlike a linear model (\figref{Fig2}). This can be attributed to the nonlinear gain saturation, which prevents arbitrary close modes from reaching threshold. In general, the stability of the laser depends on the relationship between the decay rates of the three fields, $\gamma_\perp$ for $\mathbf{P}$, $\gamma_\parallel$ for $D$, and $\kappa$ for $\mathbf{E}$, the decay rate of the cavity in the absence of gain \cite{ohtsubo2012semiconductor}. When two (or more) of these decay rates become similar, we notice a sharp reduction of $D_0$ for the onset of instability (in this case, $\gamma_\parallel \sim \kappa$).

Stability can also be studied using a multimode SALT by including the first lasing mode in the gain saturation and computing the pump threshold for a second lasing mode as a function of $k$ (inset of \figref{Fig3}(a)). In particular, this coincides with the results from the stability eigenproblem in the limit $\gamma_\parallel \rightarrow 0$. Solving~(\ref{lin-MB}) for $\gamma_\parallel \rightarrow 0$ is indeed equivalent to having $\delta D \rightarrow 0$ and $\delta X$ being a solution to SALT equation. As can be seen in the inset of \figref{Fig3}(a), the nonlinear gain saturation pushes the threshold of the arbitrary close modes ($q\rightarrow 0$) to a higher pump value compared to what is expected from a linear model. However, this multimode SALT predicts a second lasing mode that is arbitrary close to the first lasing mode, which is outside the domain of validity of SALT. Furthermore, the instability onset depends rather strongly on $\gamma_\parallel$, emphasizing the need for a full Maxwell--Bloch stability analysis.

In order to check the stability of the lasing mode close to threshold for a general system, we use perturbation theory to compute $\sigma(q, D_0)$ near $(0,D_t)$. Analytical details are shown in the supplementary material, using methods similar to those developed in \citeasnoun{liu2017symmetry}. In the case of small loss, we obtain a simple approximate condition for stability near threshold: the band curvature $\re{\frac{d^2 \omega}{dk^2}}$ and the laser detuning ($\omega_t-\omega_a$) should have the \emph{same sign} at threshold. When lasing at the band edge, this is equivalent to requiring $\omega_a$ to lie inside the band gap. 

\begin{figure}[!htb]
\includegraphics[width=\columnwidth, keepaspectratio]{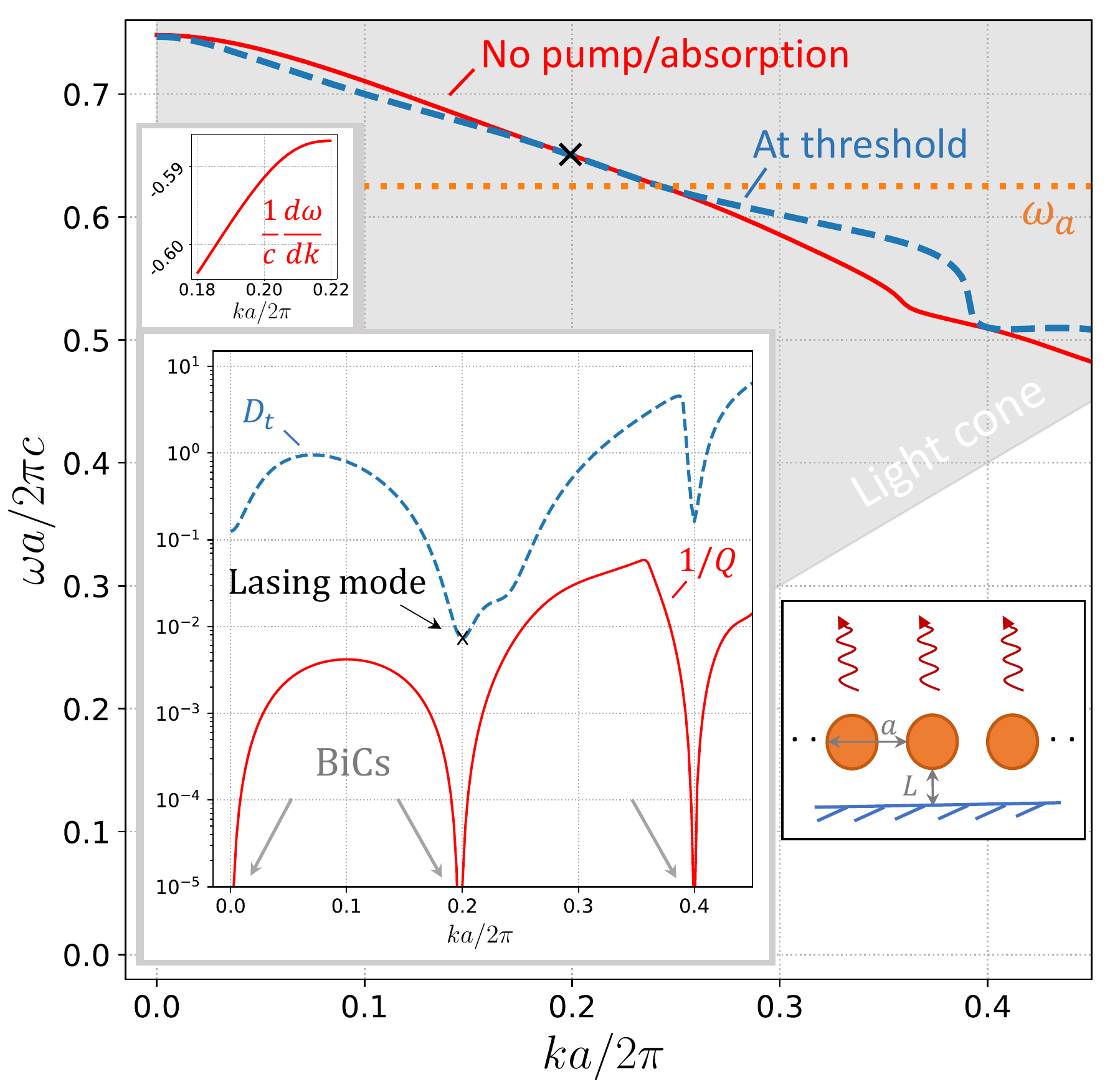}
\caption{Inset shows a 2d array of cylindrical rods with diameter $=0.7a$, $\epsilon_c=2.58$, $\sigma_c=0.001\omega_a$ and a separation $L=1.078a$ to a perfect mirror. Gain inside the rods is characterized by $\omega_aa/2\pi c=0.625$ and $\gamma_\perp a/2\pi c=0.01$. Three BiCs are shown at $ka=0, 0.4\pi, 0.8\pi$. The minimum pump at threshold $D_t$ is obtained at $ka=0.4\pi$ which is the first lasing mode. In absence of gain, the decay rate for this mode is equal to $\kappa \approx 8 \times 10^{-5} (2\pi c /a)$. Top inset shows a positive band curvature at threshold.}
\label{Fig4}
\end{figure}

We now validate the results of stability analysis against FDTD simulations~\cite{OskooiRo10, 2020arXiv} with a finite supercell and periodic boundary conditions. We initialize the simulation fields with the SALT solution plus additional noise, and analyze whether the system remains in the same steady-state at later times. Note that for a supercell with $N_\mathrm{cells}$ periods, only a finite set of values for $q$ is allowed ($=2\pi \ell/aN_\mathrm{cells}$ for $\ell = 0,\ldots,N_\mathrm{cells}-1$). \Figref{Fig3}(b) shows a perfect match between the two computations. In particular, the instability onset for the FDTD simulations corresponds to the value of the pump $D_0$ for which at least one \emph{allowed} $q$ reaches the instability region obtained from the stability eigenproblem~(\ref{instab-eigen}). Once instability is reached, a second lasing mode starts. This second lasing mode corresponds to the first $q$ that hits the instability region. However, the new lasing solution is not accurately described by two-mode SALT (\Figref{Fig3}(c)) because the small frequency difference violates the SALT assumptions (exact in the limit $\gamma_\parallel \rightarrow 0$). In particular, the inset of \figref{Fig3}(a) shows that the threshold of the multimode SALT (for $q=4\pi/50a$) does not match the actual threshold for the stability eigenproblem.  As $N_\mathrm{cells}$ increases, the second lasing frequency becomes arbitrary close to the first mode, requiring an ever-smaller $\gamma_\parallel$ for the multimode SALT approach to be viable. On the other hand, for a fixed $N_\mathrm{cells}$, the multimode SALT approach becomes increasingly accurate for smaller $\gamma_\parallel$. The two-mode regime here also exhibits a chaotic behaviour, typical in certain classes of lasers~\cite{ohtsubo2012semiconductor}. 

\begin{figure}
\includegraphics[width=\columnwidth, keepaspectratio]{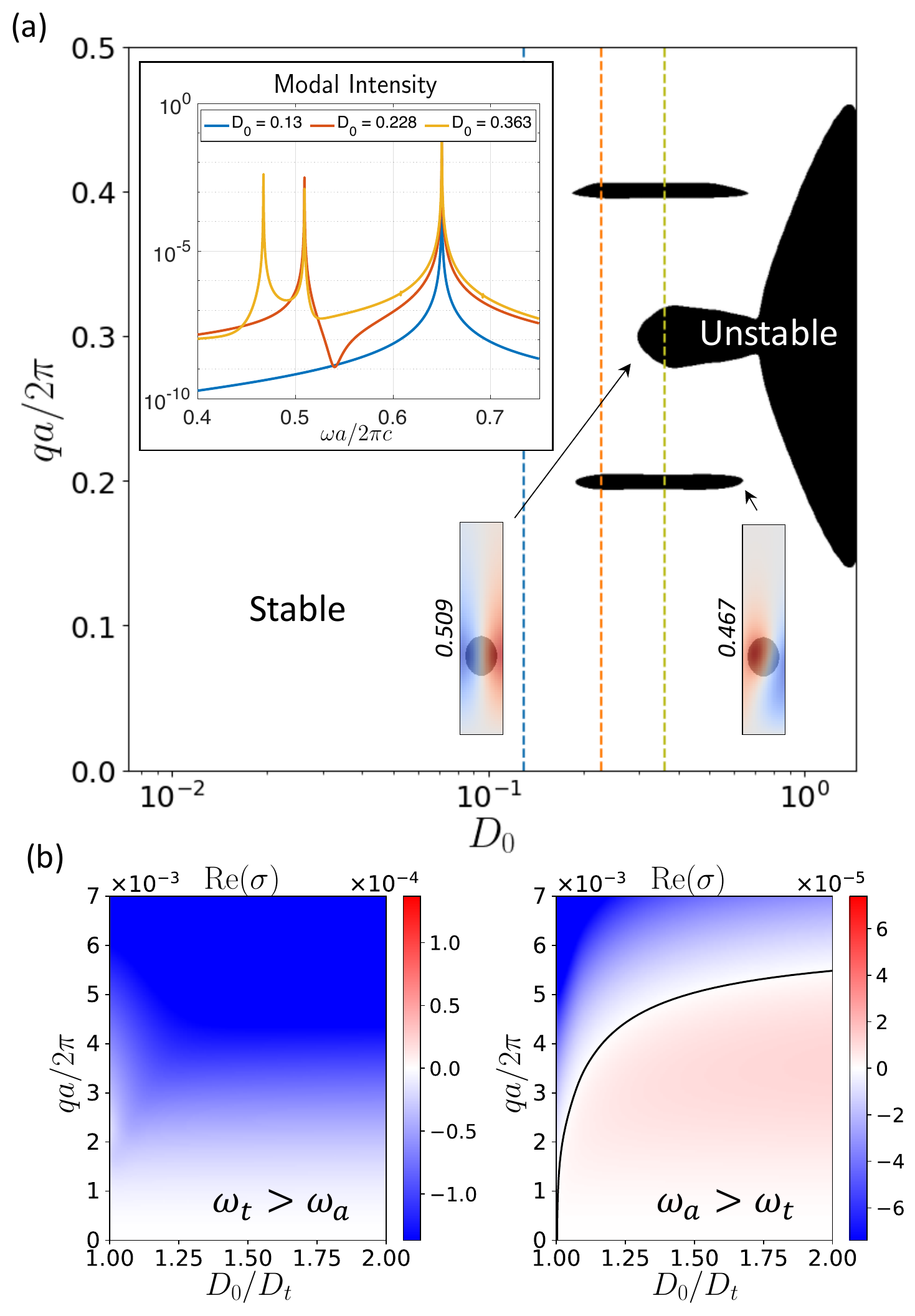}
\caption{(a) Result from stability eigenproblem. Shaded region indicates instability. Inset shows FDTD results using a supercell with 20 unit cells and periodic boundary conditions. Plots show the Fourier transform of the electric field at a point near a rod. Small insets show the eigenvectors obtained from~(\ref{instab-eigen}) along with their frequencies $\omega a /2\pi c$. They do match modes obtained in the linear regime (below threshold) at $ka=0.8\pi$ and $ka=\pi$. (b) $\text{Re}(\sigma)$ as a function of $q$ and $D_0/D_t$ for different transition frequencies $\omega_aa/2\pi c \;(=0.625,0.675)$. The threshold lasing frequency $\omega_ta/2\pi c$ is maintained at $\approx 0.65$. The system is unstable near threshold when the laser detuning ($\omega_t-\omega_a$) has opposite sign to the band curvature. Black solid line corresponds to $\re{\sigma}=0$.}
\label{Fig5}
\end{figure}

\medskip
We next consider a 2d ($E_z$-polarized) example to study the stability of a BiC lasing mode. The structure is a periodic line of surface rods placed at a distance $L$ from a perfect-metal boundary (\Figref{Fig4} inset), which is known to have multiple BiCs~\cite{hsu2013bloch}. BiCs are characterized by a quality factor $Q\rightarrow\infty$ in absence of external pump and absorption loss, as seen in the inset. As in the previous 1d example, we compute the pump threshold $D_t$ at different wavevectors $k$ and find the lasing mode corresponding to the smallest $D_t$. In this example, the first lasing mode corresponds to the BiC at $ka = 0.4\pi$, with $D_t \approx 7 \times 10^{-3}$ and a lasing frequency $\omega_ta/2\pi c \approx 0.65$. The results of the stability analysis are shown in \figref{Fig5}(a) for $\gamma_\parallel a/2\pi c=5\times10^{-3}$. We first note that the lasing mode is stable near threshold and that instability occurs at a higher pump value $D_0$ [\figref{Fig5}(b-left)]. This matches our condition for stability near threshold (positive band curvature and laser detuning). As clear from the corresponding $q$ and eigenfrequencies, instabilities at higher pump correspond to modes that become active at $ka=0.8\pi$ (BiC) and $ka=\pi$ (guided mode). A comparison between our stability results and FDTD simulations is shown in \figref{Fig5}(a-inset), where we plot the Fourier transform of the electric field at a given point outside a rod for different pump values. The number and frequencies of lasing modes match our stability computations. Finally, in order to confirm our simple stability condition, we study the same system with a larger $\omega_a$ corresponding to a negative laser detuning. As shown in \figref{Fig5}(b-right), the lasing system is indeed \emph{not} stable for any value of pump above threshold. Such instabilities may arise in very large systems (small $q$).   

\medskip

The method presented in this Letter gives a rigorous answer to the fundamental question of stable lasing in infinite periodic systems, and provides practical guidance in the form of theoretical criterion for stability. If these criteria are satisfied, the main theoretical challenges for future work are to analyze the effects of boundaries (which we expect are negligible for sufficiently large systems) and manufacturing disorder (which must eventually limit single-mode lasing).

\bigskip

See the supplementary material for analytical details of perturbation theory.

This work was supported in part by the U.S. Army Research Office through the Institute for Soldier Nanotechnologies under award W911NF-18-2-0048.

The data that supports the findings of this study are available within the article and its supplementary material.

\bibliography{main_biblio}

\begin{onecolumngrid}
\newpage


\section*{Supplementary material}

# Supplementary Material
# Is single-mode lasing possible in an infinite periodic system?

Mohammed Benzaouia,[1] Alexander Cerjan,[2] and Steven G. Johnson[3]

[1]*Department of Electrical Engineering and Computer Science,*
*Massachusetts Institute of Technology, Cambridge, MA 02139, USA.*
[2]*Department of Physics, Penn State University, State College, PA 16801, USA.*
[3]*Department of Mathematics, Massachusetts Institute of Technology, Cambridge, MA 02139, USA.*

In the main text, we showed how to apply numerical stability analysis to evaluate the stability of any lasing mode for any given system. In this supplementary material, we obtain general analytical results for the specific question of stability near lasing threshold.

In particular, we use perturbation theory to compute the stability eigenvalues $\sigma(q = q_0 + \delta k, d)$ for small $\delta k$, where $D_0 = D_t(1 + d^2)$ with $D_t$ being the pump at threshold, for points $q_0$ where $\sigma(q_0, 0) = 0$. We validate our semi-analytical results against brute-force stability eigenvalues computed as in the main text, showing excellent agreement. The perturbation theory is particularly subtle due to eigenvalue crossings that result in "critical lines" where $\sigma$ changes form, and these are also reproduced in the numerical validation. The final result is a formula that determines stability near threshold in terms of simple integrals of the threshold lasing mode. In the limit of low-loss resonances, this result further simplifies to a criterion relating band curvature to gain detuning as mentioned in the main text.

## I. PERTURBATION ANALYSIS

In all systems, we have by definition $\sigma(0, 0) = 0$. For reciprocal systems, the mode at $-k$ also reaches threshold at $D_t$ so that $\sigma(\pm 2k, 0) = 0$ [1]. Note that this last case does not have to considered when $k$ and $-k$ are separated with lattice vectors, as for example when lasing at a band edge or at the center of the Brillouin zone. We first give a detailed derivation in the case $q_0 = 0$, and then present the results for $q_0 = \pm 2k$.

The stability eigenproblem is given by $\left(A_q + B\sigma + C\sigma^2\right)U_q = 0$, where:

$$A_q = \begin{pmatrix} \Delta_{k,q}^r & -\Delta_{k,q}^i & \omega^2 & 0 & 0 \\ \Delta_{k,q}^i & \Delta_{k,q}^r & 0 & \omega^2 & 0 \\ \gamma_\perp D & 0 & \omega_a - \omega & \gamma_\perp & \gamma_\perp \mathbf{E}^r \\ 0 & \gamma_\perp D & -\gamma_\perp & \omega_a - \omega & \gamma_\perp \mathbf{E}^i \\ -\gamma_\parallel \mathbf{P}^i & \gamma_\parallel \mathbf{P}^r & \gamma_\parallel \mathbf{E}^i & -\gamma_\parallel \mathbf{E}^r & \gamma_\parallel \end{pmatrix}, B = \begin{pmatrix} -\sigma_c & -2\epsilon_c\omega & 0 & -2\omega & 0 \\ 2\epsilon_c\omega & -\sigma_c & 2\omega & 0 & 0 \\ 0 & 0 & 0 & 1 & 0 \\ 0 & 0 & -1 & 0 & 0 \\ 0 & 0 & 0 & 0 & 1 \end{pmatrix}, C = \begin{pmatrix} -\epsilon_c & 0 & -1 & 0 & 0 \\ 0 & -\epsilon_c & 0 & -1 & 0 \\ 0 & 0 & 0 & 0 & 0 \\ 0 & 0 & 0 & 0 & 0 \\ 0 & 0 & 0 & 0 & 0 \end{pmatrix}$$
(S1)

with $\Delta_{k,q}^r = -e^{-iqx}\text{Re}(\Theta_k)e^{iqx} + \epsilon_c\omega^2$, $\Delta_{k,q}^i = -e^{-iqx}\text{Im}(\Theta_k)e^{iqx} + \sigma_c\omega$, $\mathbf{E}^r = \text{Re}(\mathbf{E})$ and $\mathbf{E}^i = \text{Im}(\mathbf{E})$. For brevity of notation, we removed the subscript $k$ from $\omega_k$, $\mathbf{E}_k$, $\mathbf{P}_k$, $D_k$, but vectors still refer to the periodic part of Bloch terms. The SALT mode can be expanded in $d$, as for example done in Ref. 1. In particular, we have:

$$\omega \approx \omega_t + \omega_2 d^2, \ \mathbf{E} \approx d\frac{a\mathbf{E}_+}{\Gamma_t}, \ |a|^2 = \frac{G_D + \omega_2 H}{I}, \ \omega_2 = -\text{Im}\left(\frac{G_D}{I}\right)/\text{Im}\left(\frac{H}{I}\right)$$
(S2)

where $\mathbf{E}_+$ (resp. $\mathbf{E}_-$) is a solution to the linear SALT equation at threshold with Bloch vector $k$ (resp. $-k$). $G_D$, $I$ and $H$ are given by:

$$G_C = \int d\mathbf{x}(\epsilon_c + i\sigma_c\omega_t) \mathbf{E}_-\cdot\mathbf{E}_+, \ G_D = \int d\mathbf{x}D_t \ \mathbf{E}_-\cdot\mathbf{E}_+, \ I = \int d\mathbf{x}D_t|\mathbf{E}_+|^2\mathbf{E}_-\cdot\mathbf{E}_+, \ H = \frac{1}{\omega_t^2\Gamma_t}\frac{\partial}{\partial\omega_t}\left[\omega_t^2\left(G_C + G_D\Gamma_t\right)\right].$$
(S3)

Note that there is an arbitrary choice for the phase of $a$. To simplify some computations, we take $a\Gamma_t^*$ to be real.

Operators $A_q$, $B$ and $C$ can then be expanded in $(\delta k = q - q_0, d)$:

$$A_q \approx A_{00} + A_{01}d + A_{02}d^2 + A_{10}\delta k + A_{20}\delta k^2, \quad B \approx B_0 + B_2d^2, \quad C = C_0.$$
(S4)

As a result, eigenvalues and eigenvectors can be expanded in the same way:

$$U_q \approx \sum_{i,j\le 2} U_{ij}\delta k^i d^j, \ \sigma \approx \sum_{i,j\le 2} \sigma_{ij}\delta k^i d^j.$$
(S5)

A crucial point that we confirm later, is that $\sigma$ is *not* necessarily analytical at $(q_0, 0)$ since there is a degeneracy. So equation (S5) is not valid inside a ball around $(\delta k, d) = (0, 0)$. Instead, we have different expansion coefficients depending on the path $(\delta k, d)$.

We first consider $q_0 = 0$. The zeroth-order stability problem is equivalent to the threshold SALT equation at $k$. Because real and imaginary parts of the field are split, we have two degenerate eigenvectors $v_p$ corresponding to $\sigma_{00} = 0$, where:

$$v_p = \left( \text{Re} \left( \mathbf{e}_p^+ \right), \ \text{Im} \left( \mathbf{e}_p^+ \right), \ D_t \text{Re} \left( \Gamma_t \mathbf{e}_p^+ \right), \ D_t \text{Im} \left( \Gamma_t \mathbf{e}_p^+ \right), \ 0 \right), \tag{S6}$$

for $\mathbf{e}_{1,2}^+ = \mathbf{E}_+, \ i\mathbf{E}_+$. We also need solutions $w_p$ to the transverse problem $w_p^t A_{00} = 0$ given by:

$$w_p = \left( \text{Re} \left( \mathbf{e}_p^- \right), \ -\text{Im} \left( \mathbf{e}_p^- \right), \ \frac{\omega_t^2}{\gamma_\perp} \text{Re} \left( \Gamma_t \mathbf{e}_p^- \right), \ -\frac{\omega_t^2}{\gamma_\perp} \text{Im} \left( \Gamma_t \mathbf{e}_p^- \right), \ 0 \right), \tag{S7}$$

where $\mathbf{e}_{1,2}^- = \mathbf{E}_-, \ i\mathbf{E}_-$.

We now have $U_{00} = b_1 v_1 + b_2 v_2$, where $b_p$ are to be determined by degenerate perturbation theory. As we will see later, the coefficients $b_p$ depend on the path $(\delta k, d)$. To simplify notations, we note $\bar{M} = \left[ w_j^t M v_p \right]_{jp}$ for a given operator matrix $M$. The first order perturbation equations are given by:

$$\begin{aligned}
(\delta k) & \quad (B_0 \sigma_{10} + A_{10}) U_{00} + A_{00} U_{10} = 0 \ \rightarrow \ \bar{A}_{10} b = -\sigma_{10} \bar{B}_0 b \\
(d) & \quad (B_0 \sigma_{01} + A_{01}) U_{00} + A_{00} U_{01} = 0 \ \rightarrow \ \bar{A}_{01} b = -\sigma_{01} \bar{B}_0 b.
\end{aligned} \tag{S8}$$

It is straightforward to show that $\bar{A}_{01} = 0$, $\bar{B}_0 = -\text{Im} \left( \omega_t^2 \Gamma_t H M \right)$ and $\bar{A}_{10} = i \text{Im} \left( L M \right)$, where $M = \begin{pmatrix} 1 & i \\ i & -1 \end{pmatrix}$ and $L = -\int d\mathbf{x} \ \mathbf{E}_- \cdot \partial_q \Theta_{k+q} \mathbf{E}_+$ (in particular, $-\partial_q \Theta_{k+q} = 2i \ e^{-ikx} \nabla e^{ikx}$ for $\mathbf{E} = \text{E}_z \mathbf{z}$ waves). We then have:

$$\sigma_{01} = 0, \quad \sigma_{10} = i \frac{L}{\omega_t^2 \Gamma_t H} \text{ or } \sigma_{10} = i \left( \frac{L}{\omega_t^2 \Gamma_t H} \right)^*. \tag{S9}$$

Since 0 is a maximum of $\text{Re}[\sigma(\delta k, 0)]$, $\sigma_{01}$ is purely imaginary and the two eigenvalues are identical. So $\bar{A}_{10} + \sigma_{10} \bar{B}_0 = 0$ and $b$ is not determined by first order equations. Note that $i\sigma_{10}$ is simply the slope of $\omega(k)$ at the lasing $k$. We can also see that:

$$U_{01} = -\sum b_p g_p + \sum c_l v_l, \quad U_{10} = -\sum b_p A_{00}^{-1} (\sigma_{10} B_0 + A_{10}) v_p + \sum \tilde{c}_l v_l, \tag{S10}$$

where $g_p^5 = 2 D_t \text{Re} \left( \Gamma_t a^* \mathbf{e}_p^+ \cdot \mathbf{E}_+^* \right)$ and the first fourth components of $g_p$ are zero. $c_l$ and $\tilde{c}_l$ are arbitrary complex coefficients that will not affect our results. Note also that the fifth component of $U_{10}$ is equal to zero.

The second order perturbation equations are now given by:

$$\begin{aligned}
(\delta k d) & \quad \sigma_{11} B_0 U_{00} + (A_{10} + \sigma_{10} B_0) U_{01} + A_{01} U_{10} + A_{00} U_{11} = 0 \\
(\delta k^2) & \quad (A_{20} + \sigma_{20} B_0 + \sigma_{10}^2 C) U_{00} + (A_{10} + \sigma_{10} B_0) U_{10} + A_{00} U_{20} = 0 \\
(d^2) & \quad (A_{02} + \sigma_{02} B_0) U_{00} + A_{01} U_{01} + A_{00} U_{02} = 0.
\end{aligned} \tag{S11}$$

We start by solving the three equations independently. From results of first-order perturbation we can see that $w_j^t (A_{10} + \sigma_{10} B_0) U_{01} = 0$ and $w_j^t A_{01} U_{10} = 0$. The equation of order $\delta k d$ then gives $\sigma_{11} = 0$.

Multiplying the equation of order $\delta k^2$ by $w_j^t$ we get:

$$-\sigma_{20} \bar{B}_0 b = \left( \bar{A}_{20} + \sigma_{10}^2 \bar{C} + \bar{P} \right) b = \text{Re} \left( X M \right) b, \quad \text{where } P = (\sigma_{10} B_0 + A_{10}) A_{00}^{-1} (\sigma_{10} B_0 + A_{10}), \tag{S12}$$

where eigenvalues are simply related to the curvature of $\omega(k)$ at the lasing $k$ ($= i\sigma_{20}$):

$$\sigma_{20} = i \frac{X}{\omega_t^2 \Gamma_t H} \text{ or } \sigma_{20} = -i \left( \frac{X}{\omega_t^2 \Gamma_t H} \right)^*, \quad b = (1, \mp i). \tag{S13}$$

The degeneracy is artificially due to the separation of the real and imaginary parts of the field, so $X$ can be easily recovered from the non-degenerate perturbation theory of $\omega(k)$ in $k$. We obtain:

$$X = \int d\mathbf{x} \; \mathbf{E}_- \cdot \Box \mathbf{E}_+, \quad \Box = \partial_q^2 \Theta_{k+q} - \frac{\sigma_{10}^2}{2} \partial^2 G + (i\partial_q \Theta_{k+q} + \sigma_{10} \partial G)(-\Theta_k + G)^{-1}(i\partial_q \Theta_{k+q} + \sigma_{10} \partial G),$$

$$G(\omega_t) = \omega_t^2 \left[ \epsilon_c + i\frac{\sigma_c}{\omega_t} + D_t \Gamma(\omega_t) \right] \text{ and } \partial_q^2 \Theta_{k+q} = -I \text{ for } \mathbf{E} = \mathrm{E}_z \mathbf{z} \text{ waves.}$$

Finally, multiplying the equation of order $d^2$ by $w_j^t$ we get (using $a\Gamma_t^* = a^* \Gamma_t$):

$$- \sigma_{02} \bar{B}_0 b = \left( \bar{A}_{02} - \bar{Q} \right) b, \quad \text{with } \bar{Q} = \left[ w_j^t A_{01} g_p \right]_{jp} = \mathrm{Re} \left[ \omega_t^2 \Gamma_t |a|^2 I \left( M' + M \right) \right] \text{ and } \bar{A}_{02} = 0, \quad (S15)$$

where $M' = \begin{pmatrix} 1 & -i \\ i & 1 \end{pmatrix}$. The eigenvalues are then given by:

$$\sigma_{02} = 0, \; b = (0, 1) \quad \text{or} \quad \sigma_{02} = 2|a|^2 \mathrm{Im} \left( \frac{I}{H} \right), \; b = (-\mathrm{Im}[I/H], \mathrm{Re}[I/H]). \quad (S16)$$

We see that we obtain different eigenvectors in (S13) and (S16). This means that the expansion in (S5) depends on the path $(\delta k, d)$. If $d = o(\delta k)$, the expansion is determined by (S13); while it is determined by (S16) if $\delta k = o(d)$. A critical behaviour is obtained along the linse $\delta k = \alpha d$ for which the second order term is given by $\sigma_2 d^2$ and the three equations in (S11) have to be combined. In this case, the second order perturbation eigenproblem becomes:

$$- \sigma_2 \bar{B}_0 b = \left[ \alpha^2 \mathrm{Re} \left( XM \right) - \bar{Q} \right] b, \quad (S17)$$

and the eigenvalues are given by:

$$\sigma_2 = \mathrm{Im} \left( \alpha^2 \theta + \eta_I \right) \pm \sqrt{|\eta_I|^2 - \left[ \mathrm{Re} \left( \alpha^2 \theta + \eta_I \right) \right]^2}, \quad \theta = -\frac{X}{\omega_t^2 \Gamma_t H}, \; \eta_I = |a|^2 \frac{I}{H}. \quad (S18)$$

Note that $\theta$ is simply the band curvature at threshold ($\omega(k) \approx \omega_t + i\sigma_{10}\delta k + \theta\delta k^2$).

The presence of the square root function clearly shows the non-analyticity of $\sigma$. In particular, the there is an eigenvalue crossing for $\alpha_c^2 = (-\mathrm{Re}(\eta_I) \pm |\eta_I|)/\mathrm{Re}(\theta)$. The stability condition ($\sigma_2 \leq 0$) can also be immediately retrieved:

$$\boxed{\alpha_s^2 = -2\mathrm{Re}(\eta_I/\theta) \leq 0.} \quad (S19)$$

We can simplify the stability condition in the limit of small loss. In this case, $H \approx 2\omega_t \int \epsilon_c \mathbf{E}_- \cdot \mathbf{E}_+/\Gamma_t$, $\mathbf{E}_- \approx \mathbf{E}_+^*$ and $\mathrm{Im}(\theta) \approx 0$. The stability condition $\mathrm{Re}(\eta_I) \mathrm{Re}(\theta) + \mathrm{Im}(\eta_I) \mathrm{Im}(\theta) \geq 0$ becomes equivalent to:

$$\boxed{\mathrm{Re}(\theta) (\omega_t - \omega_a) \gtrsim 0.} \quad (S20)$$

This means that the sign of the detuning $(\omega_t - \omega_a)$ should be the same as the sign of the band curvature ($\mathrm{Re}[\theta]$). For example, when lasing at a bandedge, this means that $\omega_a$ should be inside the bandgap.

As mentioned in the beginning of the section, in the case of degenerate lasing, the previous analysis should also be carried out at $q_0 = -2k$ (or eqivalently at $2k$). (Note that we are not considering the special case of a degeneracy that comes for a wavevector other than $-k$. However, this situation can be studied in a similar way by computing a perturbation expansion of $\sigma$ around multiple adequate $q_0$s.) It is easy to see that the solutions of the zeroth order problem $A_{-2k}U_{00} = 0$ are related to solutions of SALT at $k \pm 2k$. Two separate cases should then be considered.

*a. $ka = \pi/2$:* In this case, the problems at $-k$ and $3k$ are equivalent (separated by a lattice vector) and the zeroth order problem is degenerate. The eigenvectors are given by:

$$v_p = e^{i\pi x/a} \left( \mathrm{Re} \left( e^{-i\pi x/a} \mathbf{e}_p^- \right), \; \mathrm{Im} \left( e^{-i\pi x/a} \mathbf{e}_p^- \right), \; D_t \mathrm{Re} \left( e^{-i\pi x/a} \Gamma_t \mathbf{e}_p^- \right), \; D_t \mathrm{Im} \left( e^{-i\pi x/a} \Gamma_t \mathbf{e}_p^- \right), \; 0 \right), \quad (S21)$$

while solutions of the transverse problem become:

$$w_p = e^{-i\pi x/a} \left( \mathrm{Re} \left( e^{i\pi x/a} \mathbf{e}_p^+ \right), \; -\mathrm{Im} \left( e^{i\pi x/a} \mathbf{e}_p^+ \right), \; \frac{\omega_t^2}{\gamma_\perp} \mathrm{Re} \left( e^{i\pi x/a} \Gamma_t \mathbf{e}_p^+ \right), \; -\frac{\omega_t^2}{\gamma_\perp} \mathrm{Im} \left( e^{i\pi x/a} \Gamma_t \mathbf{e}_p^+ \right), \; 0 \right). \quad (S22)$$

We now have $g_p^5 = 2D_t e^{i\pi x/a} \mathrm{Re}\left(\Gamma_t a^* e^{-i\pi x/a} \mathbf{e}_p^- \cdot \mathbf{E}_+^*\right)$ and $\bar{Q} = \mathrm{Re}\left[\omega_t^2 \Gamma_t |a|^2 (KM' + JM)\right]$, where:

$$J = \int d\mathbf{x}\, D_t (\mathbf{E}_+^* \cdot \mathbf{E}_-)(\mathbf{E}_+ \cdot \mathbf{E}_+) \text{ and } K = \int d\mathbf{x}\, e^{2i\pi x/a} D_t (\mathbf{E}_-^* \cdot \mathbf{E}_+)(\mathbf{E}_+ \cdot \mathbf{E}_+). \tag{S23}$$

We can then obtain the eigenvalues of the problem (S17) for $\delta k = q + 2k = \alpha d$:

$$\sigma_2 = \mathrm{Im}\left(\alpha^2 \theta + \eta_J\right) \pm \sqrt{|\eta_K|^2 - \left[\mathrm{Re}\left(\alpha^2 \theta + \eta_J\right)\right]^2}, \quad \eta_J = |a|^2 \frac{J}{H}, \ \eta_K = |a|^2 \frac{K}{H}. \tag{S24}$$

The stability condition is now equivalent to:

$$\boxed{\alpha_s^2 = -\mathrm{Re}\left(\frac{\eta_J}{\theta}\right) + \sqrt{\left|\frac{\eta_K}{\theta}\right|^2 - \left|\frac{\eta_J}{\theta}\right|^2 + \mathrm{Re}\left(\frac{\eta_J}{\theta}\right)^2}} \text{ non-real or real negative.} \tag{S25}$$

*b. $ka \neq \pi/2$:* In this case, the problems at $-k$ and $3k$ are different, and only $-k$ has a solution. The zeroth order problem for $q_0 = -2k$ is now *not* degenerate and eigenvectors are given by:

$$v = (1, \ -i, \ D_t \Gamma_t, \ -i D_t \Gamma_t, \ 0)\, \mathbf{E}_-, \ w = \left(1, \ i, \ \frac{\omega_t^2 \Gamma_t}{\gamma_\perp}, \ i\frac{\omega_t^2 \Gamma_t}{\gamma_\perp}, \ 0\right) \mathbf{E}_+. \tag{S26}$$

The dimension of our problem is now one and we have $g^5 = 2D_t \Gamma_t a^* \mathbf{E}_+^* \cdot \mathbf{E}_-$, $\bar{B}_0 = 2i\omega_t^2 \Gamma_t H$, $A_{20} = 2X$ and $\bar{Q} = 2\omega_t^2 \Gamma_t |a|^2 J$. The unique eigenvalue of (S17) is now equal to:

$$\sigma_2 = -i(\theta \alpha^2 + \eta_J). \tag{S27}$$

This simply means that there is no eigenvalue crossing and that the expansion of $\sigma$ does not depend on the path $(\delta k, d)$. Note that $\sigma_2^*$ is also an eigenvalue around $q_0 = 2k$ (which is is simply due to the facts that our operators $A, B$ and $C$ are real as indicated in the main text). The stability condition is immediately given by:

$$\boxed{\mathrm{Im}\left(\eta_J\right) \leq 0,} \tag{S28}$$

since we already have $\mathrm{Im}\left(\theta\right) \leq 0$ ($\mathrm{Im}[\omega(k)]$ has a maximum at $k$). Note that this stability condition is equivalent to having a stable lasing close to threshold for the *single* unit-cell problem.

Finally, some useful points to mention:

- We have $\eta_I = G_D/H + \omega_2$. It is also straightforward to use perturbation theory to show that $\omega_2^l = -G_D/H$ where $\omega_2^l$ is the slope (in $D_0/D_t - 1$) of the eigenfrequency of the linear problem at the threshold *without* gain saturation ($\omega^l \approx \omega_t + \omega_2^l(D_0/D_t - 1)$). By definition, threshold should be reached from below the real axis, so $\mathrm{Im}\left(\omega_2^l\right) \geq 0$. Since $\omega_2$ is real, we conclude that $\mathrm{Im}\left(\eta_I\right) = -\mathrm{Im}\left(\omega_2^l\right) \leq 0$. This means that $\sigma_{02} \leq 0$ and that the single unit-cell lasing problem is always stable near threshold in absence of degeneracy.

- For TM waves ($\mathbf{E} = \mathrm{E}_z \mathbf{z}$), we have $I = J$. This means that $\mathrm{Im}\left(\eta_J\right) \leq 0$ and that the single unit-cell lasing problem is also stable in the degenerate case when $k \neq \pi/2$. This is an analytical proof for part of the stability result conjectured in Ref. 1. Note that $k = \pi/2$ is equivalent to the condition $n = 4\ell$ in Ref. 1.

- For TM waves and $k \neq \pi/2$, we conclude that $\sigma_2 \leq 0$ when expanding around $-2k$. So the stability is only determined by the expansion around 0 ($-\mathrm{Re}\left(\eta_I/\theta\right) \leq 0$.)

## II. NUMERICAL VALIDATION

Here, we present a numerical validation of the analytical perturbation-theory results discussed in the previous section.

Figure S1 shows results for the 1d structure studied in the main text. Figs. S2–S3 are for the same structure, but with $\omega_a$ lying below the lasing band edge, outside the bandgap, leading to instability near threshold as predicted above. In both cases, the numerical simulations show near-perfect agreement with the analytical results.

Figures S5–S4 show results for the 2d structures presented in the main text with a positive and negative laser detuning, respectively. Again, numerical simulations are in agreement with the analytical results.

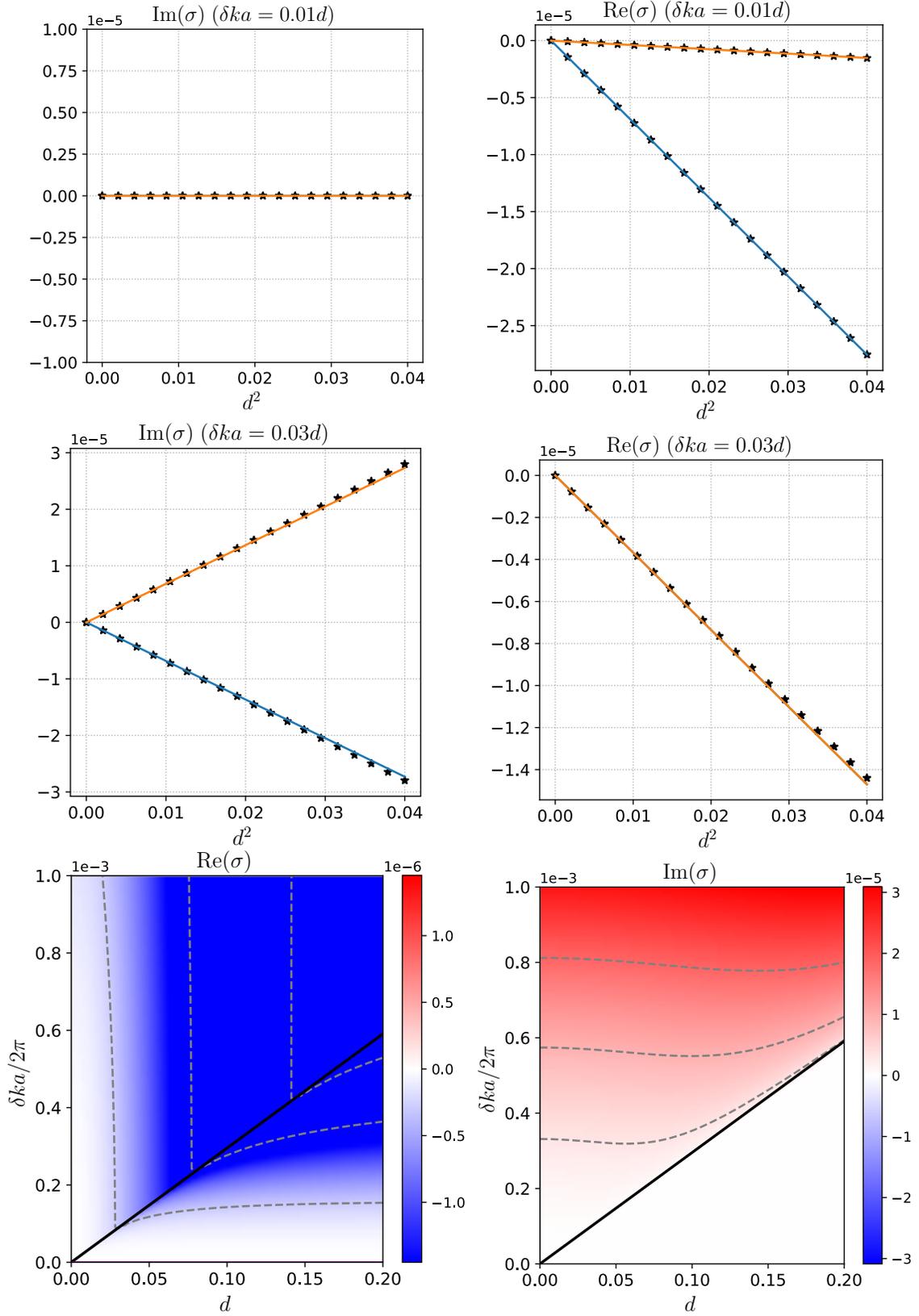

FIG. S1. Same 1d structure in the main text. Numerical simulation (stars and dashed contour lines) are in agreement with analytical results (solid lines). Since the lasing mode is at a bandedge, we have $\sigma_{10} = 0$. Black line corresponds to $\delta ka = \alpha_c d$ and represents the line of eigenvalue crossing (transition from two real to two complex conjugate eigenvalues). $\alpha_c \approx 0.018$ and $\alpha_s^2 \approx -4.2 \times 10^{-4}$.

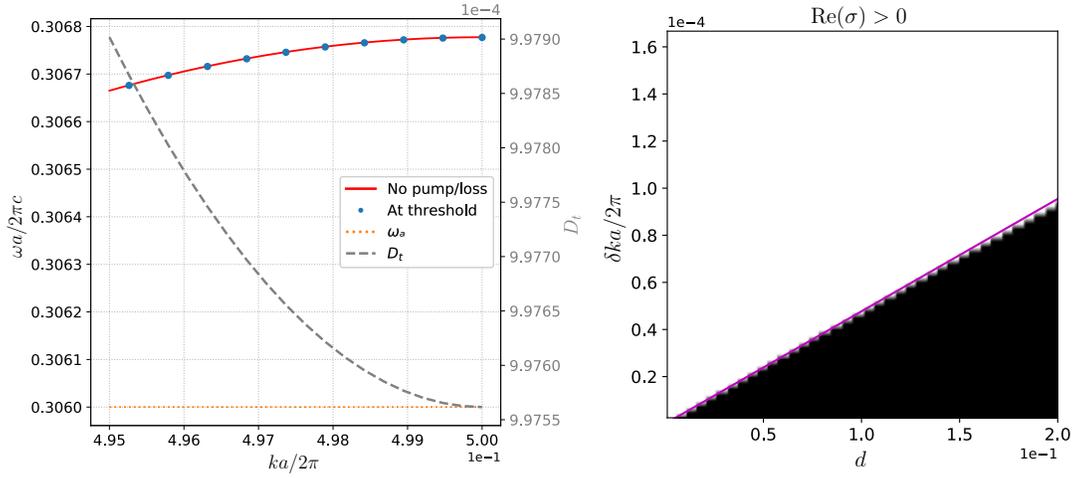

FIG. S2. Same 1d structure in the main text but with $\omega_a a/2\pi c = 0.306$ and $\gamma_\perp/2\pi c = 0.08$. The lasing mode is still at the band edge but the laser detuning ($\omega_t - \omega_a$) is now positive.

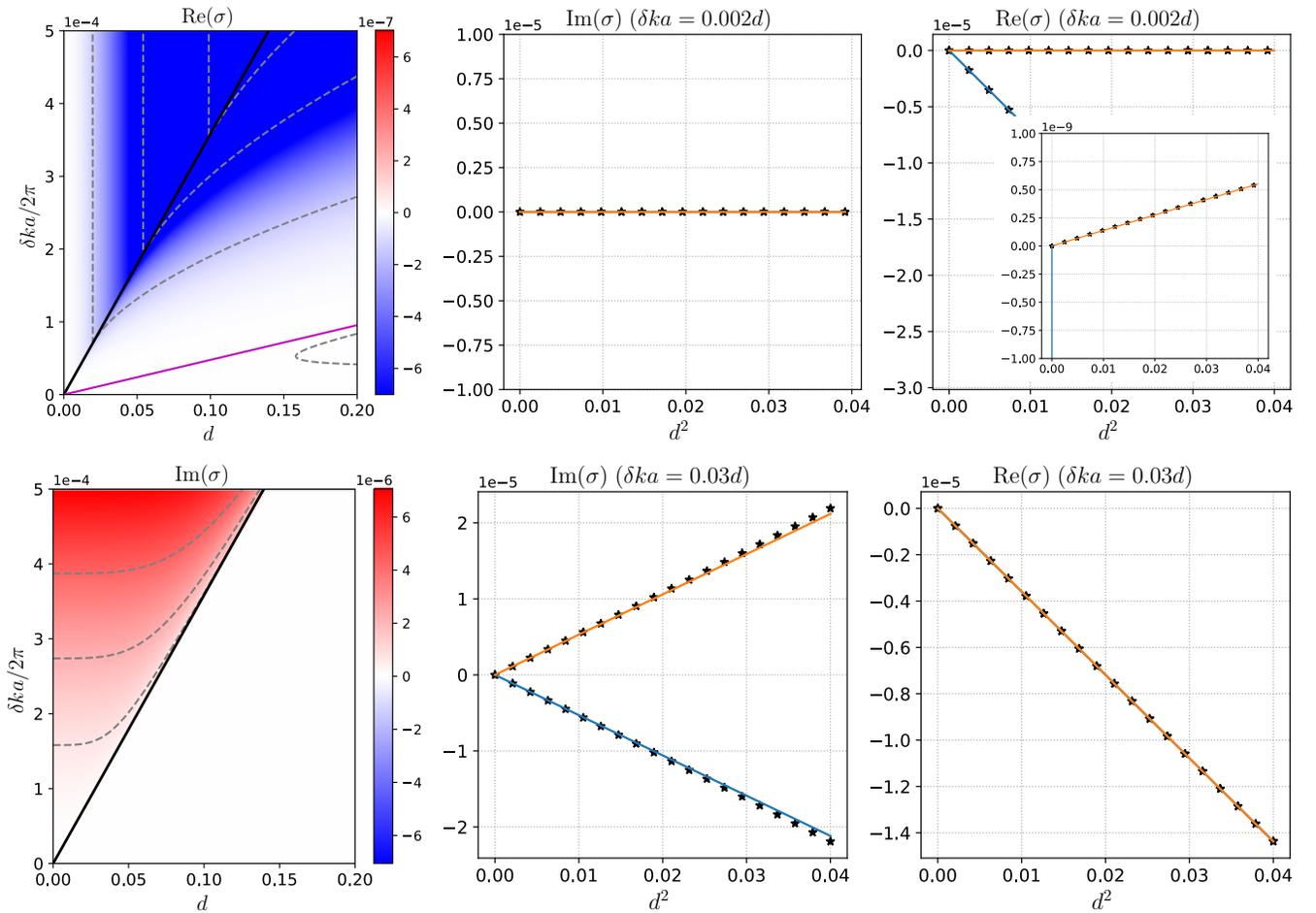

FIG. S3. Same 1d structure studied in the main text but with $\omega_a a/2\pi c = 0.306$ and $\gamma_\perp/2\pi c = 0.08$. Numerical simulation (stars and dashed contour lines) are in agreement with analytical results (solid lines). Black line corresponds to $\delta ka = \alpha_c d$ and represents the line of eigenvalue crossing (transition from two real to two complex conjugate eigenvalues). Magenta solid line corresponds to $\delta k = \alpha_s d$ from analytical perturbation results and matches $\text{Re}(\sigma) = 0$ from numerical simulation. $\alpha_c \approx 0.022$ and $\alpha_s \approx 3 \times 10^{-3}$.

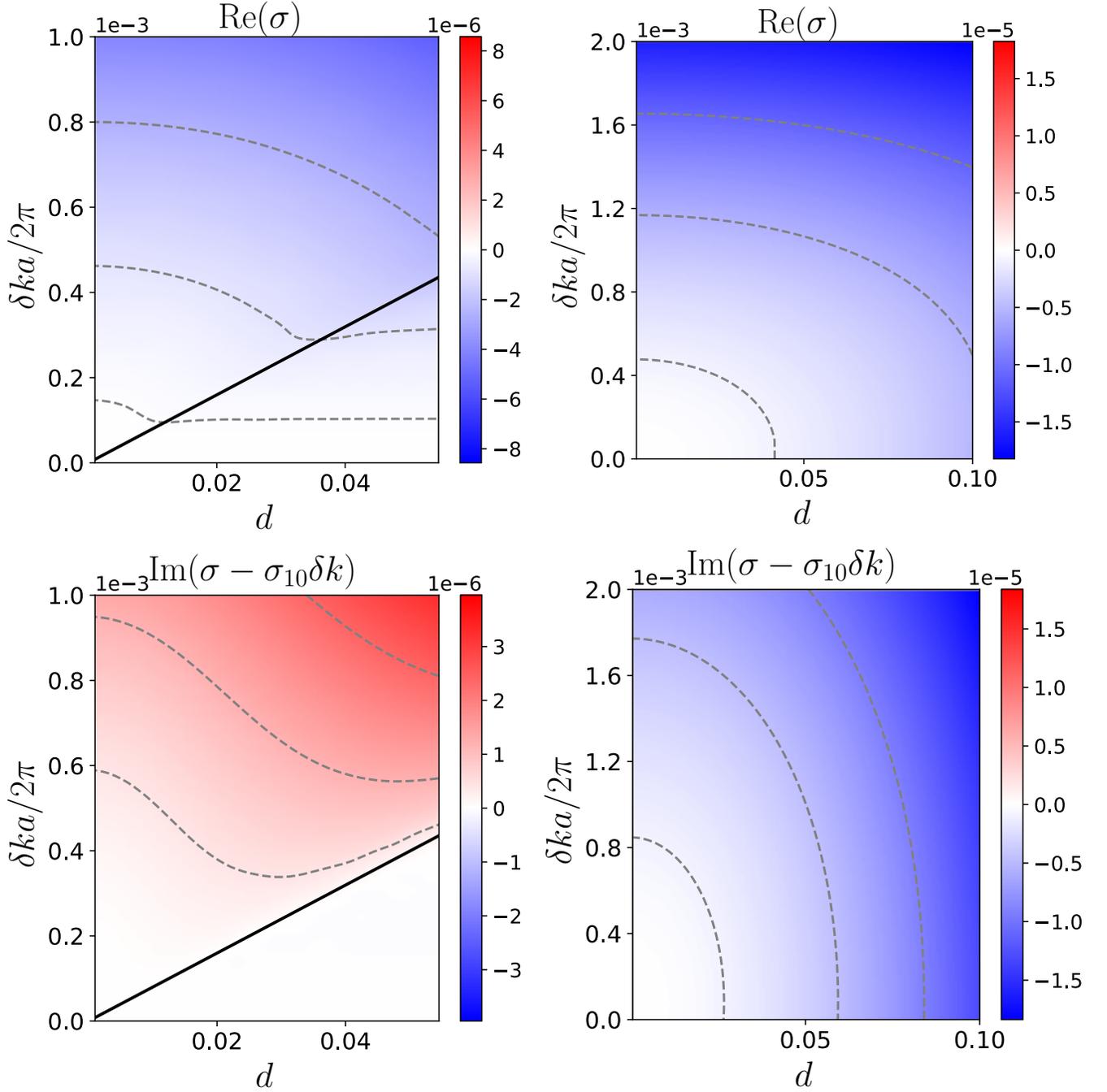

FIG. S4. Same 2d structure in the main text with $\omega_a a/2\pi c = 0.625$ and $\omega_t a/2\pi c \approx 0.65$. Left: $q_0 = 0$. Right: $q_0 = -2k$. Contour lines (dashed) are from numerical simulation. Black solid line corresponds to $\delta k = \alpha_c d$ from analytical perturbation results and represents the line of eigenvalue crossing (transition of $\sigma - \sigma_{10}\delta k$ from two real to two complex conjugate eigenvalues) when expanding around $q_0 = 0$. The analytical line matches results of numerical simulation. Expansion around $-2k$ does not show a critical line in agreement with perturbation theory (case $ka \neq \pi/2$). We have $\alpha_c \approx 0.05$, $\alpha_s^2 \approx -0.018$ and $\sigma_{10} \approx 0.59i$ when expanding around $q_0 = 0$ (opposite sign for $i\sigma_{10}$ when expanding around $q_0 = -2k$).

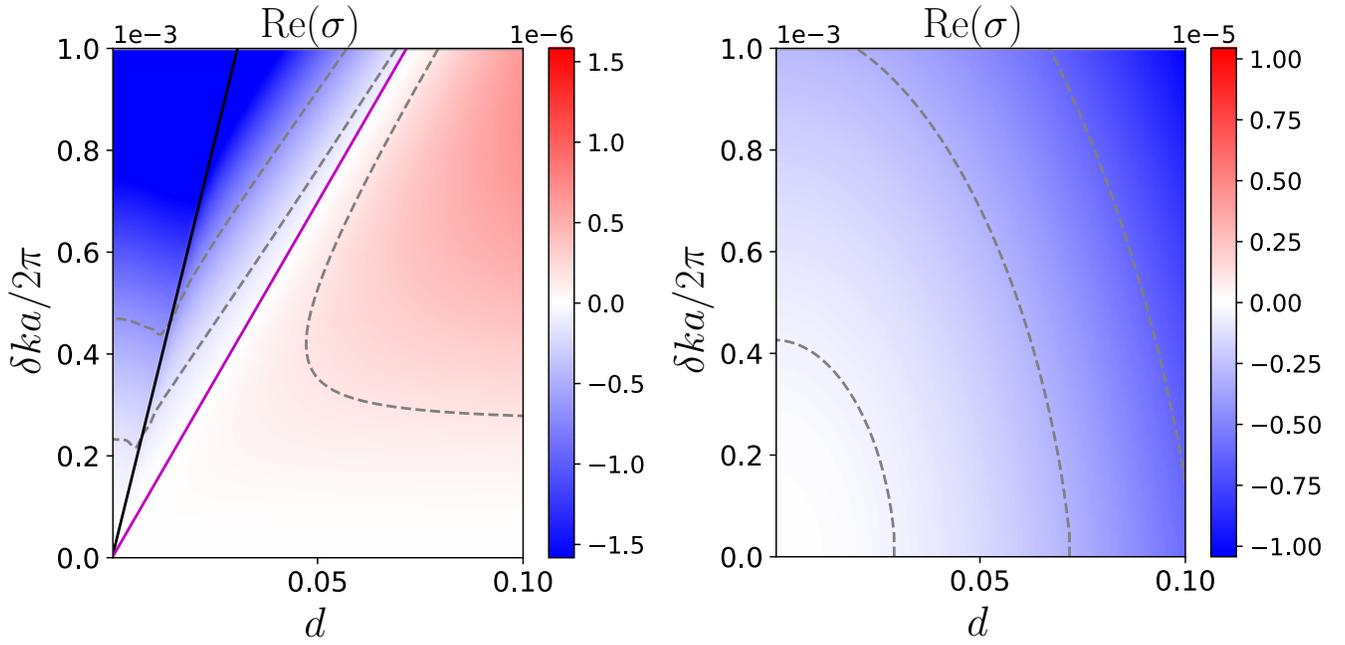

FIG. S5. Same 2d structure in the main text with $\omega_a a/2\pi c = 0.675$. Left: $q_0 = 0$. Right: $q_0 = -2k$. The lasing mode is slightly shifted to $ka/2\pi \approx 0.1944$ but still with $\omega_t a/2\pi c \approx 0.65$. Contour lines (dashed) are from numerical simulation. Black solid line corresponds to $\delta k = \alpha_c d$ and magenta solid line corresponds to $\delta k = \alpha_s d$ from analytical perturbation results when expanding around $q_0 = 0$. Majenta line (analytical) matches $\mathrm{Re}(\sigma) = 0$ from numerical simulation. Expansion around $-2k$ does not show a critical line in agreement with perturbation theory (case $ka \neq \pi/2$). We have $\alpha_c \approx 0.21$, $\alpha_s \approx 0.088$ and $\sigma_{10} \approx 0.59i$ when expanding around $q_0 = 0$ (opposite sign for $i\sigma_{10}$ when expanding around $q_0 = -2k$).


\thispagestyle{empty}
\end{onecolumngrid}

\end{document}